# Tearing Graphene Sheets From Adhesive Substrates Produces Tapered Nanoribbons

Dipanjan Sen[1], Kostya S. Novoselov[2,*], Pedro M. Reis[3], and Markus J. Buehler[1,†]

[1] *Laboratory for Atomistic and Molecular Mechanics, Department of Civil & Environmental Engineering, Massachusetts Institute of Technology,*
*77 Massachusetts Avenue, Cambridge, MA 02139, USA*

[2] *School of Physics & Astronomy, University of Manchester, Oxford Road, Manchester, M13 9PL, UK*

[3] *Department of Mathematics, Massachusetts Institute of Technology,*
*77 Massachusetts Avenue, Cambridge, MA 02139, USA*

[*] kostya@manchester.ac.uk

[†] mbuehler@mit.edu

**Graphene is a truly two-dimensional atomic crystal with exceptional electronic and mechanical properties. Whereas conventional bulk and thin film materials have been studied extensively, the key mechanical properties of graphene, such as tearing and cracking, remain unknown, partly due to its two-dimensional nature and ultimate single-atom-layer thickness, which result in the breakdown of conventional material models. By combining first-principles ReaxFF molecular dynamics and experimental studies, a bottom-up investigation of the tearing of graphene sheets from adhesive substrates is reported, including the observation of the formation of tapered graphene nanoribbons. Through a careful analysis of the underlying molecular rupture mechanisms, it is shown that the resulting nanoribbon geometry is controlled by both the graphene–substrate adhesion energy and by the number of torn graphene layers. By considering graphene as a model material for a broader class of two-dimensional atomic crystals, these results provide fundamental insights into the tearing and cracking mechanisms of highly confined nanomaterials.**

## 1. Introduction

Studies of the mechanical behavior of crystals have thus far focused primarily on three-dimensional (3D) bulk crystalline materials, such as metals or ceramics, thin-film materials, quasi-1D structures (e.g., carbon nanotubes), as well as quasi-0D materials (e.g., buckyballs). However, the mechanical properties of 2D crystals at an intermediate level of dimensionality remain largely unexplored, and their fundamental properties such as mechanical stability, defect evolution, and fracture mechanisms remain unknown. The first experimental observation of isolated graphene,[1,2] single freestanding sheets of graphite, has generated significant interest in the electronics,[3–6] mechanics,[7,8] and nanoscience communities. This is because graphene provides, for the first time, access to a truly 2D atomic crystal with remarkable material properties. The unique properties of graphene include high electron mobility and tunable bandgaps, depending on the width of the graphene nanoribbon,[9,10] making graphene a promising material for innovative electronic applications. Furthermore, the extreme in-plane stiffness of graphene,[8] along with the absence of structural defects, lead to extremely high fracture strengths close to the theoretical limit. These properties render graphene a viable material for the development of novel composite materials.[11–13]

However, the applicability of conventional models of bulk or thin-film materials to describe the mechanical properties of graphene under extreme conditions is questionable because the thickness of graphene reaches the ultimate limit of one atomic layer, representing the thinnest of all films and thus a severe departure from the structural makeup that underlies models applied to conventional materials. On the other hand, the elucidation of the mechanical properties of graphene presents an exciting opportunity that could have a profound impact in advancing our fundamental understanding of 2D atomic crystals. For example, many conventional methods of graphene manufacture involve mechanical processes that induce large deformations or breaking of covalent bonds Thus, a widely used technique for exfoliation of few-and mono-layer graphene is based on tearing off graphene layers using adhesive tape.[1,2] Another technique is based on epitaxial deposition of graphene on a crystal, followed by removal of the underlying substrate.[14,15] In both approaches, the tearing of graphene sheets from graphite or other substrates is used to obtain freestanding sheets or narrow ribbons through the application of mechanical loading, as shown schematically in Figure 1a. Other methods to manufacture graphene nanoribbons[9,10,16–19] and graphene oxide nanoribbons[20] also involve large deformations and fracture in graphene sheets. Despite the broad application of these graphene manufacturing approaches, the fundamental mechanisms of large deformation and tearing mechanisms of graphene remain unknown.

To provide insight into the mechanisms of graphene under tear loading, we carried out a series of experiments that reveal that ribbon-like structures are formed, as displayed in Figures 1b–d. Notably, the graphene ribbons produced using this technique consistently result in a tapered geometry, where the angles of tear are found to depend on the substrate used as well as on the number of layers torn. While the tapered ribbons observed in our tearing experiments are reminiscent of macroscopic triangular tears obtained during peeling of adhesive films (such as Scotch tape) from substrates,[21] the formation of tapered graphene ribbons remains unexplained as of today. Specifically, the dimensions and adhesion strengths in graphene are greatly different from the conditions described in macro-scale tearing experiments and associated models.[21] As a result, conventional continuum theory cannot be applied to explain the tearing phenomena in graphene. The mechanisms behind the formation of tapered nanostructures, the specific angles observed, the effect of the underlying discrete atomic lattice, the influence of the 2D nature of the graphene sheet, as well as the influence of geometric parameters such as the number of graphene layers torn remain unknown.

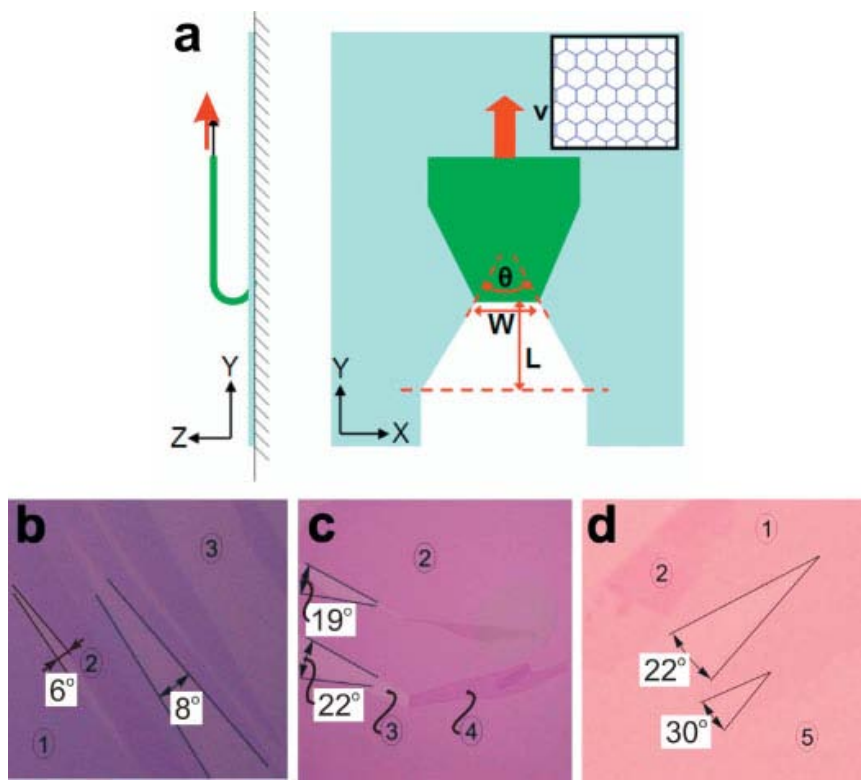

Figure 1. Overall geometry of the tearing setup and experimental results of peeling. a) Schematic diagram of the setup for the tearing studies of graphene: side and top views; the inset shows the sheet orientation. An initial flap of 80 Å in width is cut in the sheet, folded back, and moved at a constant speed. Typical graphene flakes obtained by micromechanical cleavage:[9,10] b) monolayer graphene on $SiO_2$ (total width of the panel: 150 μm); c) bilayer graphene on $SiO_2$ (total width of the panel: 150 μm); d) monolayer graphene on PMMA (total width of the panel: 20 μm). Circled numbers: 1 = monolayer graphene, 2 = bilayer graphene, 3 = $SiO_2$, 4 = folded bilayer graphene (2+2 layers), 5 = PMMA. In the experimental studies, a thick (1–100 nm) flake was deposited first, which was then torn off, leaving monolayer or bilayer graphene with characteristic wedge-like tears. Note that, although it is difficult to obtain proper statistics of the tearing angle (due to low reproducibility of the experiments, which depend on many parameters, like cleanness of the substrate, tearing direction, etc.), the general trend is represented well in the figure. Specifically, the tearing angles for bilayer graphene are generally larger than for monolayers – compare the angles in panels (b) and (c) – and also larger for graphene torn off from relatively highly adhesive substrates (such as PMMA) compared with low-adhesion substrates (such as $SiO_2$) – compare the angles in panel (d) with those in panels (b) and (c).

Earlier studies on graphene mechanics have focused on the absence of extensive plastic deformation under loading.[8,22,23] However, the generation and role of defects on the fracture mechanism of this 2D material, as well as the effect of the presence of a substrate, have not been investigated. Furthermore, little is known about the loading conditions that lead to specific edge types produced by tearing and fracture of graphene. Understanding these issues is, however, critical for the development of manufacturing approaches to produce functional graphene nanostructures with defined geometries and edge characteristics. A bottom-up atomistic approach is crucial if the experimental observations shown in Figure 1b–d are to be explained as it can provide a fundamental perspective of the material properties of graphene that incorporates chemical effects associated with bond stretching and breaking, as well as the capacity to reveal the tearing mechanisms of materials at much larger scales of tens of nanometers.[24–29] Here we report a systematic atomistic simulation study using the first-principles-based reactive force field ReaxFF[30] to elucidate the fundamental fracture mechanical properties of graphene under tearing loading, as shown in Figure 1a, mimicking the loading conditions used in our experiments (Figure 1b–d). The first-principles-based ReaxFF force field has been shown to provide an accurate account of the chemical and mechanical behavior of hydrocarbons, graphite, diamond, carbon nanotubes, and other carbon nanostructures, while it is capable of treating hundreds of thousands of atoms with near quantum-chemical accuracy.

In our simulations, two parallel crack notches were initially created in a single large graphene sheet (with a size of 200Å×175Å) adhered to a substrate and the edge of the flap connecting the tears was lifted, bent back, and pulled at a constant velocity in the y direction to mimic the experimental tearing load. We focused our study on the dependence of the tearing process on, firstly, the adhesion strength between graphene and a substrate and, secondly, on the number of layers torn. These layers are arranged in the graphite Bernal A–B stacking. For all conditions considered here, we present a detailed analysis of the atomistic-level mechanisms and their respective relationship to the macroscopically observed tear angle. Further details of the experimental and computational protocols are provided in the Experimental Section.

## 2. Results and Discussion

We begin our analysis by focusing on the shape of the tearing path under varying adhesion strength by carrying out a series of computational experiments. Generally, we find that the width of the ridge joining the two crack tips at either end of the tear narrows as tearing progresses through the material for all adhesive strengths considered. As shown in Figure 2, the narrowing of the torn graphene ridge leads to the formation of tapered tears, in direct agreement with the findings revealed in our experiments (Figure 1b–d). We find that the shape of the resulting tear is a function of the strength of adhesion. Figure 2a–c shows snapshots of the tearing geometry for increasing adhesion strength. We find that the tear edges are composed of a combination of armchair and zigzag edges, reflecting the discrete nature of the underlying 2D atomic lattice. An average angle of tearing (marked as $\theta$ in Figure 1a) can be measured by plotting the width of the tear ridge, $W$, as a function of the length torn, $L$, and by fitting a linear relationship, as presented in Figure 2d.

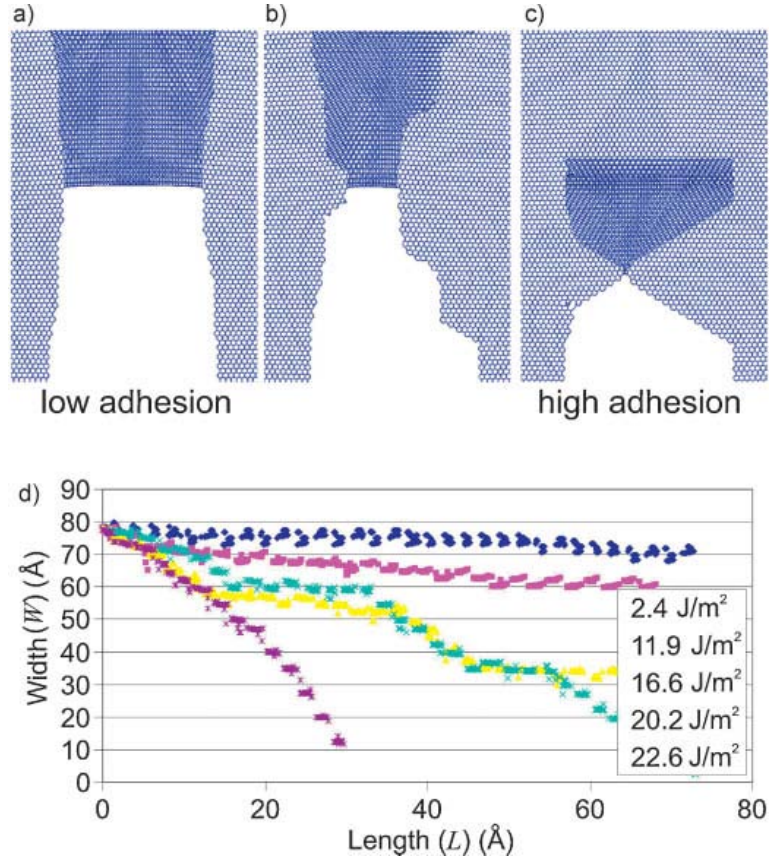

Figure 2. Molecular dynamics simulation results of graphene tearing. a–c) Structure of torn nanoribbons at different adhesion strengths (substrate not shown for clarity). d) Variation of the width of the ridge ($W$) with the length of the flap from the initial position ($L$) for different adhesion strengths (see legend). Linear fits are made to all curves to obtain the average rate of change of the width with respect to height and thus the average angle of tearing.

The observed narrowing of the torn section at all adhesive strengths is also predicted by continuum theory, where it has been developed for and applied to tearing of macroscopic adhesive films.[21] This existing model is based on considering an energy balance between the release of bending elastic energy (due to the advancement of the crack tips) and the narrowing of the tear ridge. By formalizing this energy balance, this macro-scale continuum theory predicts that

$$\sin\frac{\theta}{2} \propto \frac{(D\tau)^{1/2}}{\gamma t} \qquad (1)$$

where $\theta$ is the tearing angle, $D$ is the out-of-plane bending modulus of the thin sheet, $\tau$ is the adhesion strength per unit area, $\gamma$ is the work of fracture per unit area, and $t$ is the thickness of the torn film. Based on its derivation, this model applies to systems with large adhesion energies ($\tau \times$ width of flap) compared to their fracture strength ($\gamma t$). We now test the predictions of this continuum theory of tearing of elastic sheets against the behavior of graphene as obtained from our molecular dynamics simulations. We emphasize that the macro-scale continuum theory predicts a uniform angle of the torn section, thereby yielding triangular tears with straight edges. Also, the tear edges are predicted to be mirror symmetrical with respect to the plane formed by the tearing direction ($y$) and the out-of-plane direction ($z$) (Figure 1a). However, while we observe in both our experiments and simulations that tapering does indeed occur, we find that the tear edges are neither straight (the tearing angle varies along the tear path), nor mirror symmetric, in disagreement with the macroscopic continuum theory.

We now compare the scaling behavior predicted from the continuum theory with molecular simulation results. Figure 3a shows a plot of the sine of half of the angle of tearing against the square root of the adhesion strength, $\tau$. According to Equation 1, continuum theory predicts a linear scaling between these two quantities, regardless of the adhesion strength. However, in contrast to this prediction, our molecular simulations show that there exist two regions of distinct scaling behavior, depending on the adhesion strength. The first regime is well described by a linear fit (following the behavior predicted by Equation 1) $\sin(\theta/2) \propto \sqrt{\tau}$. In contrast, the second regime shows a rather different scaling, where the sine of the angle scales with the adhesion strength as $\sin(\theta) \propto \tau^2$.

To distinguish the two regimes, the adhesions strengths are henceforth referred to as ''low'' (where $\tau = 2$–$8$ J m$^{-2}$) and ''high'' (where $\tau > 8$ J m$^{-2}$), depending on which regime in Figure 3a they lie on. An analysis of the atomistic stress distribution in the adhered sheet close to the tear ridge shows that the out-of-plane shear stress ($\sigma_{ZX}$) and the tensile stress in the y direction ($\sigma_{YY}$) are both increased under increasing adhesion (Figure 4). At low adhesion, the direction of fracture is close to the direction of maximum out-of-plane shear. However, at high adhesion, the tensile-stress field ahead of the bent section and in the flap, and thus the elastic stretching of the sheet, play a major role in reducing the flap width and therefore dictate crack progression. The continuum theory[21] developed based on the tearing of macroscopic films, however, does not account for this effect of stretching-energy storage ahead of the tear ridge and in the torn flap. To account for the effect of the elastic stretching energy in the system, we have extended the theory (details of the derivation are provided in the Supporting Information). The new theory predicts the scaling behavior as

$$\sin\frac{\theta}{2} \propto \tau^2 \qquad (2)$$

in excellent agreement with our computational results. The results depicted in Figure 3a confirm that the stretching energy in the graphene sheet becomes the dominant contribution to the elastic energy of the system for single-sheet graphene under large adhesion, as opposed to macroscopic adhesive films, where this contribution from stretching can be neglected. This illustrates an important distinction between macro-scale tearing and tearing of a 2D atomic crystal. Notably, the framework of continuum theory can still be applied to provide a good model of the behavior of this 2D nanoscale system once the stretching contribution is properly accounted for.

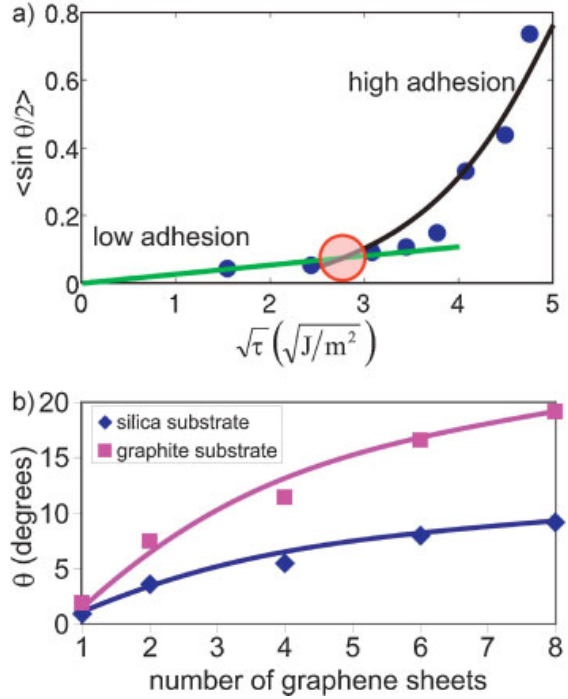

Figure 3. Analysis of tearing angle as a function of adhesion strength and number of graphene sheets torn. a) A plot of the sine of half of the angle of tearing against the square root of the adhesion energy. We identify two regimes of different behavior: a linear one with slope of $0.026 \pm 0.003$ m J$^{-0.5}$ and a scaling of half of the angle of tearing as the square of the adhesion energy, with a prefactor of $0.0012 \pm 0.0001$ m$^4$ J$^{-2}$. The crossover point is highlighted by a red circle and the two regimes are referred to in the text as ''low'' and ''high'' adhesion regimes, respectively. b) Predicted angle of tear for graphene adhesion on silica and graphite surfaces (adhesion energies of 0.08 and 0.345 J m$^{-2}$, respectively) as a function of the number of sheets torn off simultaneously.

Extrapolating the low-adhesion part of the curve in Figure 3a to the graphite-silica ($\tau = 0.08$ J m$^{-2}$) and graphene– graphene interlayer ($\tau = 0.345$ J m$^{-2}$) adhesion energy[34] as used in the experiment, we obtain tearing angles of 0.92° and 1.9° for the tearing of a single graphene sheets from silica and graphite surfaces, respectively. According to the continuum analysis, the angle of tear is proportional to the bending modulus of the film and inversely proportional to the thickness. We further extend our predictions to multisheet tearing of graphene (assuming that the low-adhesion continuum formulation of Equation 1 holds) by calculating the bending modulus, $D$, as a function of the number of layers of graphene (results shown in Table 1). Based on these calculations, Figure 3b shows the angle of tear as a function of the number of graphene sheets torn simultaneously from silica and graphite surfaces. This result suggests that a strong variation of the angle of tear is expected depending on the number of layers of graphene being torn. Although it is difficult to put forward a detailed statistical analysis of the tearing angle in the experimental studies (due to the relatively low reproducibility of the experiments, which depend on many parameters, such as the cleanliness of the substrate, tearing direction, etc.), the general trend predicted by molecular simulations is well represented in Figure 1b–d. Specifically, the tearing angles for bilayer graphene are generally larger than for monolayers, as can be confirmed by comparing the angles in Figure 1b and c. Moreover, the angles are larger for graphene on relatively high adhesivity substrates, such as poly(methyl methacrylate) (PMMA), compared with low adhesivity substrates (such as $SiO_2$), as can be seen by comparing the angles shown in Figure 1b and c with those shown in Figure 1d.

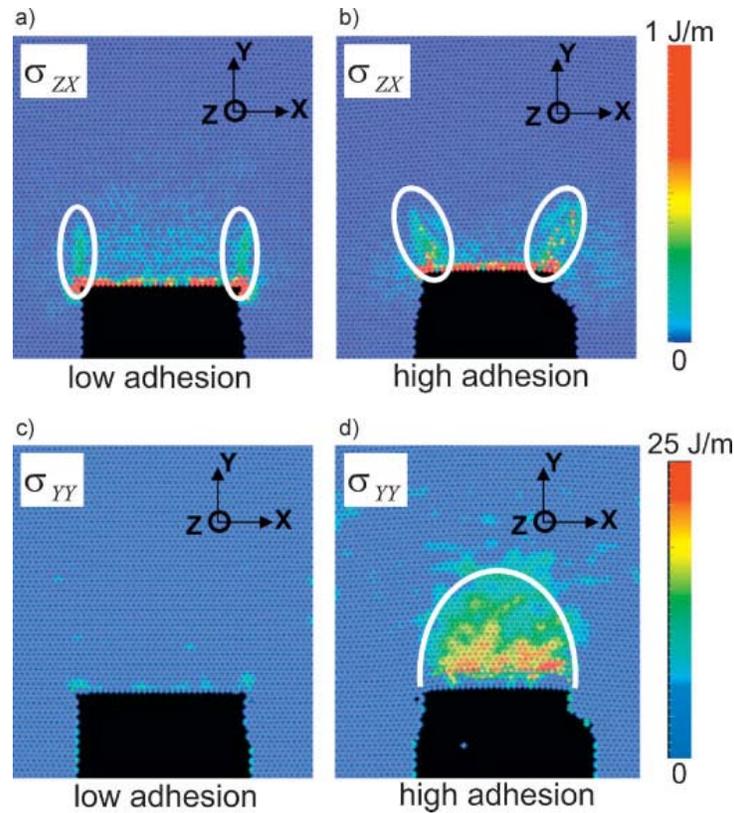

Figure 4. Atomistic-level virial stress distribution in the graphene sheet. a, b) Out-of-plane shear stress, $\sigma_{ZX}$, for adhesion strengths of 2.4 and 20.2 J m$^{-2}$ for low and high adhesion, respectively. c, d) In-plane tensile stress in y direction, $\sigma_{YY}$, for adhesion strengths of 2.4 and 20.2 J m$^{-2}$ for low and high adhesion, respectively. The results show that larger adhesion strengths lead to much larger tensile stresses in the sheet ahead of the crack bend. We also see a larger stretching energy stored in the torn flap. The release of this elastic stretching energy becomes an important parameter in dictating the crack propagation direction.

We now focus on the molecular mechanisms of tearing by analyzing the atomistic-simulation trajectories. Figure 5 shows that, at low adhesion strength, the narrowing of the tear section is achieved through small surface steps in a zigzag orientation. These alternate with portions of more extended propagation in the tearing direction: the armchair edge along the y direction (hereafter referred to as the first armchair).[35] At high adhesion strength, the narrowing is achieved through a combination of zigzag steps and a second type of armchair surface, orientated at 60° with respect to the first armchair (hereafter referred to as the second armchair).

Furthermore, we find that, at low adhesion strengths, the tear surfaces consist of long straight armchair sections parallel to the loading direction alternated by zigzag steps, which extend over the width of one or two hexagonal carbon rings (shown in Figure 5a and b). These zigzag steps are responsible for the narrowing of the tearing ridge by making the two crack tips move towards each other. The average lengths of the first armchairs (i.e., the portions of straight propagation along the y direction) between the small zigzag steps decreases as the adhesion strength increases, leading to a faster narrowing of the torn sheet. The tear direction switches easily from armchair to zigzag and back by locally choosing one of two carbon–carbon bonds to break.

A second type of fractured surface appears for higher adhesion strengths. This is the second armchair surface (shown in Figure 5c and d). The two armchair surfaces are oriented at 60° to each other. The second armchair surface sections are small in length at the beginning but increase in size with increasing adhesion strength until they are comparable or larger than the first armchair sections. The size of the zigzag sections of the tear surface remains about the same, reaching the width of 3–4 hexagonal carbon rings at the maximum. The local angle of tear is seen to increase rapidly with adhesion strength, in agreement with the experimental results shown in Figure 1b–d. We further find that, in this case, tear edges are highly asymmetrical between the left and right surfaces.

Table 1. Bending modulus, *D*, calculated as a function of the number of graphene layers. For two or more layers, the bending modulus versus the cube of the layer height yield a linear fit, in agreement with linear elastic theory of thin plates.

| Number of graphene layers | Bending modulus $D$ [eV] |
| --- | --- |
| 1 | $2.1 \pm 0.1$ |
| 2 | $130 \pm 4$ |
| 4 | $1\,199 \pm 18$ |
| 8 | $13\,523 \pm 1\,037$ |

The transition from the first armchair sections to the second ones involves the rotation of the tear surface by a large angle. As shown in Figure 6, these change-of-direction events are always accompanied by carbon-carbon bond rotations, leading to the formation of adjacent pentagon-heptagon rings (called the formation of 5–7-defect clusters) at the crack tip. The 5–7 defects change the local stress field, with the next bonds to be broken being on the curved torn section, leading to changing of the tear path by a large angle. Interestingly, 5–7 Stone–Wales defects have also been shown to be important in the deformation of carbon nanotubes.[36–38] Their formation is an Arrhenius process that is temperature or stress driven.[39] In our case, the appearance of 5–7 defects is stress driven and appears increasingly more frequently at larger adhesions due to the higher local stresses at the crack tip (as shown in the stress field plots depicted in Figure 4b and d). The presence of these defects allows the crack to bend sharply, thereby effectively reducing the width of the tearing ridge.

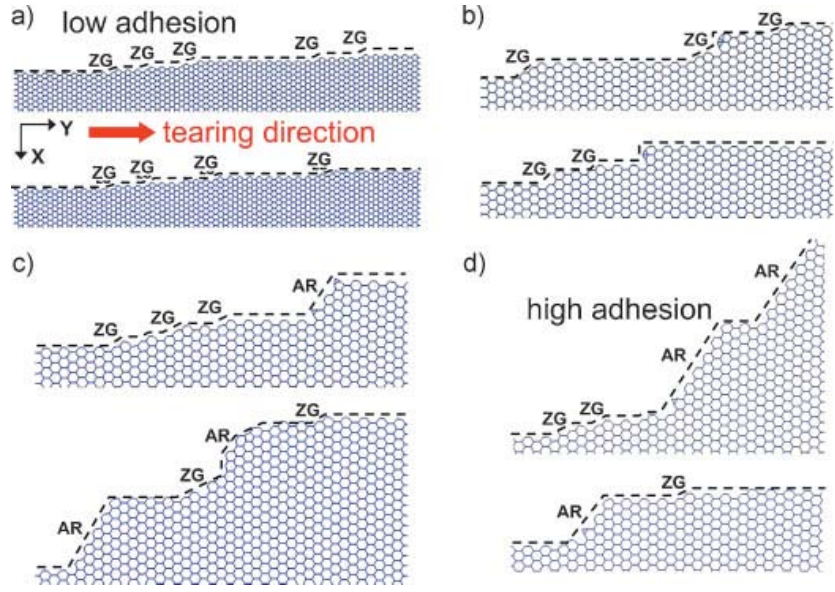

Figure 5. Atomistic tearing mechanisms for different adhesion strengths and edge geometries for the two edges formed as the tear moves forward. The two subfigures in each panel represent the shape of the two crack lips formed on either side of the tearing ridge. Results are shown for adhesion strengths of a) 2.4, b) 14.3, c) 16.7, and d) 20.2 J m$^{-2}$. At lower adhesion strengths ((a) and (b)), the fracture surface largely consists of first-armchair edges (unlabeled, horizontal dotted lines) with small zigzag ledges (marked as ZG) that provide the geometric means that reduce the width of the torn sheet. At higher adhesion strengths ((c) and (d)), the second armchair edge begins to form (marked as AR) and takes over large sections of the torn sheet.

## 3. Conclusions

In summary, in both experimental studies and atomistic simulations, we find that tearing of graphene sheets leads to tapered ribbons and the comparison of molecular simulation and experimental analysis shows good agreement. The controlling parameters that define the geometry of the torn graphene nanoribbons are the adhesion strength to the underlying substrate and the number of graphene sheets that are torn off simultaneously (Figures 1–3). We find that, at the atomistic level, the tapering is not uniform and is found to be rather strongly influenced by the discreteness of the atomic structure of the graphene lattice. The edges of the torn sections consist of a sequence of armchair and zigzag structures, with the former being more prominent, as shown in Figure 5.

Our molecular simulations and the continuum theory predict a linear relation between the sine of half of the angle of tearing and the square root of the adhesion strength at lower adhesion strengths (Equation 1). At higher adhesion strengths, the graphene tears become highly nonsymmetrical and the linear relationship breaks down. We find that this is due to an increasing contribution to the elastic energy in the system from stretching energy stored in the sheet (Figure 4), resulting in a change in the energy balance.[21] By taking the stretching energy into consideration (see the Supporting Information for details on the theory), our new continuum model accurately reflects the different regimes observed in the simulations. Specifically, the model predicts a linear relationship between the sine of half the angle of tearing and the square of the adhesion strength at large adhesion strengths (Equation 2). Macroscopic adhesive films feature a bending modulus- to-elastic-modulus ratio that scales as $t^3$, where $t$ is the thickness of a film. In contrast, single-layer graphene has a bending-modulus-to-elastic-modulus ratio that is much lower than predicted by conventional continuum theory. This effect, owing to the extreme thinness of the material, leads to an increased contribution of stretching energy to the tearing process, a phenomenon not found in macroscopic experiments of tearing elastic films from adhesive surfaces.[21] Our studies thereby provide fundamental insight into tearing mechanisms specific to the behavior of ultrathin 2D atomic crystals, which could have implications for a variety of other recently discovered 2D atomic crystals, such as graphane.[40]

As shown in Figure 1b–d, the tapered tears are observed experimentally in the peeling of few graphene layers from silica substrates. The angle of tearing measured in these experiments is in the approximate range of 1.5–11°, which is consistent with the predictions from our atomistic simulations (Figure 3b). The presence of local roughness on the underlying substrate,[41] which would lead to wrinkling in the graphene sheet, is expected to have a small effect on the tearing angles as long as the roughness is of the order of a few Å . This is expected from the small change in adhesion energy at these roughness features. In addition to the experiments reported in this article, tapered nanoribbons consisting of 1–3 layers have also been produced by breaking off from larger sheets during sonication,[10] with a similar range of tearing angles from ≈2–14°, in agreement with the simulation results. This observed variation in angle suggests that a different number of sheets have been torn off in these experiments; however, further quantitative analysis will be required to investigate this issue in more depth. Based on the results reported here, it may be possible to achieve controlled width variation and edge structures of the graphene ribbons by varying the number of sheets torn off and the type of substrate onto which the graphene is deposited. Furthermore, more controlled experimental studies of the peeling off of graphene could be carried out to provide a more quantitative comparison between the experimental results and simulation predictions. The varying width tapered graphene tears may also feature interesting band gap variations along their length, which could be studied in future work in the context of molecular electronics applications. Further, the tears show 5–7 defects at the edges when changing from one armchair edge to another, whose effect on the electronic properties could be of potential interest and a subject of future investigations.

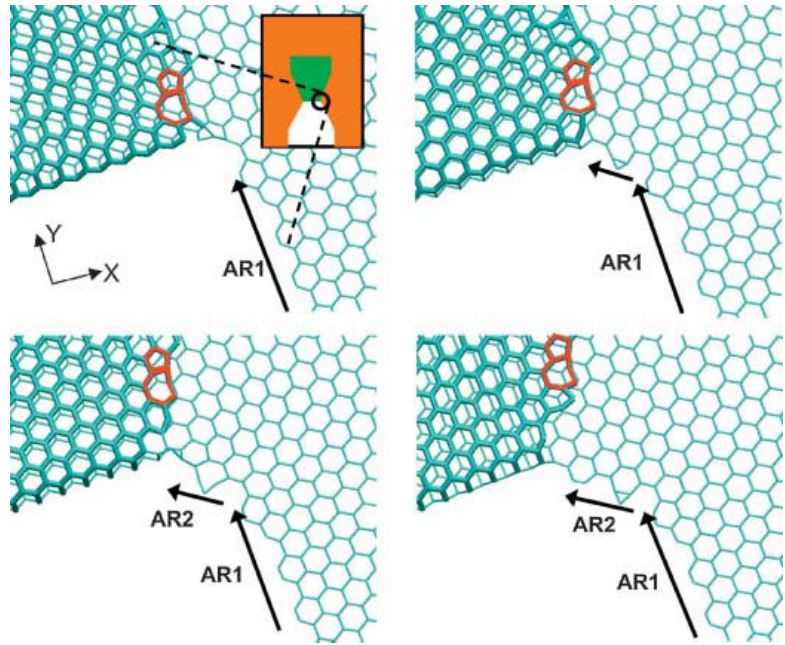

Figure 6. Role of the 5-7 defect in changing the tear direction from the first armchair (AR1) to the second armchair (AR2) edge configuration. Here, the out-of-plane flap is highlighted in bold lines, while the original sheet is represented by thinner lines. The 5-7 defects, outlined in red, are always present in transitions of the edge from the first armchair to the second armchair. The presence of these defects changes the local strain field, allowing the crack to bend sharply, thus allowing a large reduction in the width of the torn section. We find that the defect forms consistently during the change of direction of the crack tip and is carried over onto the flap.

## 4. Experimental Section

*Experimental studies*: Graphene samples were prepared by micromechanical cleavage of natural graphite. Thick flakes of graphite (with 1–100-nm thickness) were then deposited on the surface of $SiO_2$ or PMMA by pressing adhesive tape with cleaved graphite on it. Successive peeling of the thick flakes by means of a fresh adhesive tape leaves mono- and bi-layer graphene flakes with characteristic wedge tears, as shown in Figure 1b–d. High-resolution optical and scanning electron microscopy images were taken and the tearing angles were measured through digital-image processing.

*Molecular-simulation model*: We applied the first-principles-based ReaxFF reactive force field,[30] which closely retains the accuracy of quantum mechanics even for bond-breaking events. ReaxFF accurately describes mechanical properties at small and large strains and simulates bond breaking in reaction pathways. The ReaxFF parameters were determined solely by fitting to quantum mechanical data of carbon chemistry.[30] Other reactive force-fields have also been used in recent studies of the mechanics of graphene.[42–44] The six-fold rotational symmetry of the basal plane of graphene layers leads to isotropic elastic and bending moduli in the plane.[45] Since graphene is a 2D material, a nominal height of the sheet, equal to the interlayer spacing for graphite, was used to calculate elastic properties. The equilibrium sheet spacing for graphite sheets is predicted to be 3.35 Å by ReaxFF calculations. Assuming this distance for the nominal height of a single graphene sheet, the in-plane Young's modulus for graphene is calculated to be 1.01 TPa. Table 1 shows the bending modulus for multiple graphene sheets as predicted by ReaxFF. These compare well to other experimental[8,46] and simulation studies.[47,48]

*General simulation setup*: Graphene has two preferred directions of tear and edges: the armchair and zigzag orientations. Graphene nanoribbon edges have been shown to possess a combination of both armchair and zigzag configurations.[49] ReaxFF calculations for the unrelaxed surface energies for these two edges are close, 2.35 eV Å$^{-1}$ for the zigzag and 2.19 eV Å$^{-1}$ for the armchair. The armchair edge is thus lower in energy than the zigzag, in agreement with ab initio[50] and experimental[51] results. In the tearing mode considered here, the armchair configuration can be achieved by breaking every third bond of a hexagon and the zigzag by breaking every fourth (opposite) bond of a unit hexagon. To investigate the effect of adhesion to a substrate on the tearing mechanics of graphene, we introduced an adhesion potential based on a Lennard-Jones (LJ) function to model graphene–substrate interactions of different strengths. An atomically flat adhesive substrate was considered in our simulations. The adhesion strength was varied from ≈2–23 J m$^{-2}$.

Bending-modulus calculation: The isotropic bending modulus was determined by 1D pure bending experiments using molecular statics. To calculate the bending modulus of N layers of graphene, a rectangular sheet of the N layers was bent into a section of a cylinder with constant radius of curvature throughout the basal plane. The neutral plane for pure bending is parallel to the N layers and passes through the centroid of the bending cross section. The edges of the bent sheet were kept fixed and the bulk of the sheet was allowed to relax to an energy minimum. Energy minimization of the system was performed using a conjugate gradient algorithm with an energy-convergence criterion implemented in the LAMMPS code. The bending modulus was calculated by fitting the energy-curvature data to the following expression:[52]

$$E = \frac{1}{2}D\kappa^2 \qquad (3)$$

where E is the system strain energy per unit basal plane area, D is the bending modulus, and k is the beam curvature.

*Atomistic stress calculation*: The atomistic stress was calculated close to the crack tip by using the virial stress formulation.[53]

*Simulation approach*: The graphene adhesion to a surface was modeled by interaction with a LJ 9–3 wall. LJ wall potentials have been previously used in the literature to model molecular interactions with a graphite surface.[54,55] The energy of the LJ well was scaled to model different adhesion strengths for varying substrates. All tearing tests were modeled by molecular dynamics simulations performed in the microcanonical ensemble (carried out at a low temperature of 10 K to prevent large thermal vibrations). Temperature control was achieved using a Berendsen thermostat and the time step used was 1 fs. A graphene sheet of size 200 Å × 175 Å was adhered to a semi-infinite LJ 9–3 wall. The system consisted of approximately 13 000 fully reactive atoms. An initial strip of finite width was cut from the sheet and folded back, as shown in Figure 1a. The end of this strip was then displaced along the pulling direction at a pulling velocity of 1 Å ps$^{-1}$. Additional details are provided in the Supporting Information.

**Acknowledgements**

We acknowledge support from DARPA (grant number HR0011-08-1-0067) and ARO (grant number W911NF-06-1-0291). KSN is grateful to the Royal Society and the European Research

Council (programs ''Ideas'', call: ERC-2007-StG and ''New and Emerging Science and Technology,'' project ''Structural Information of Biological Molecules at Atomic Resolution'') for financial support.

*Supporting Information*

# Tearing Graphene Sheets From Adhesive Substrates Produces Tapered Nanoribbons

Dipanjan Sen[1], Kostya S. Novoselov[2,*], Pedro M. Reis[3], and Markus J. Buehler[1,†]

[1] *Laboratory for Atomistic and Molecular Mechanics, Department of Civil & Environmental Engineering, Massachusetts Institute of Technology, 77 Massachusetts Avenue, Cambridge, MA 02139, USA*

[2] *School of Physics & Astronomy, University of Manchester, Oxford Road, Manchester, M13 9PL, UK*

[3] *Department of Mathematics, Massachusetts Institute of Technology, 77 Massachusetts Avenue, Cambridge, MA 02139, USA*

[*] kostya@manchester.ac.uk

[†] mbuehler@mit.edu

**Model for elastic-stretching dominated tearing mechanics**

From the molecular dynamics simulations results presented in Figure 3a of the main text (plot of the sine of half of the tearing angle, $\sin(\theta/2)$, versus square root of the adhesion strength, $\sqrt{\tau}$), we conclude that there are two regimes in the tearing process of the graphene sheets. Moreover, the calculated atomistic stress fields in the adhered sheet close to the tear ridge shown in Figure 4 in the main text, suggest an increasing contribution of elastic stretching energy to the system energy at higher adhesion strengths. This indicates the existence of the two regimes correspond to bending-energy-dominated (at low adhesion strengths, $2 < \tau < 8$ J m$^{-2}$) and stretching-energy-dominated (at high adhesion strengths, $\tau > 8$ J m$^{-2}$) tearing.

Whereas the regime where bending energy dominates has been previously considered in the literature,[21] the corresponding theory for the stretching dominated tearing has still to be developed. The previous continuum theory[21] accounts for the release of elastic bending energy in the fold joining the two tears as the tears move forward. The bending energy is released by the simultaneous advancement of the crack tips in the Y direction and their approximation in the X direction, thereby reducing the width of the ridge. In this Supplementary Information we modify the previous model,[21] to account for the new loading and boundary conditions under which elastic stretching energy in the flap and adhered sheet become large in comparison to the bending energy and its contribution can no longer be neglected.

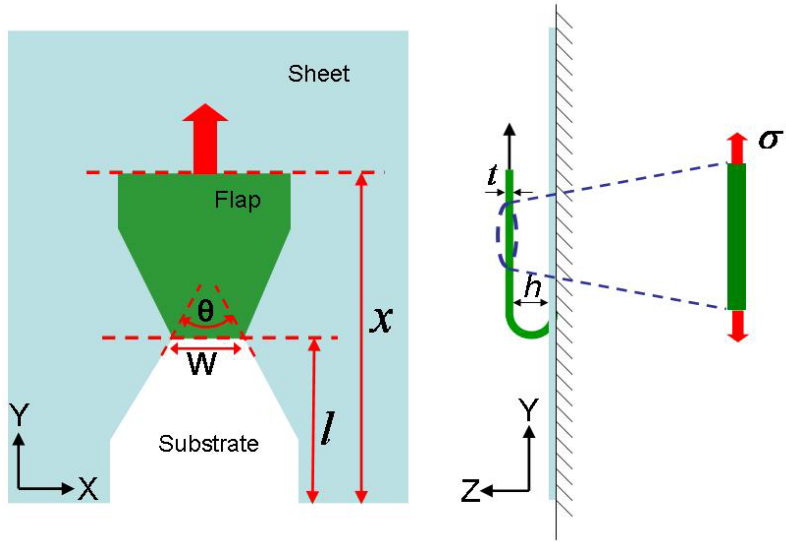

Figure S1. A schematic drawing showing the setup for the tearing studies of graphene. The plot shows a top view and side view. An initial flap is cut in the sheet, folded back and moved at a constant speed. The thickness of the sheet is t. The distance h gives the height of the fold. W is the width of the fold at the adhesion edge, l and x are the distances from the bottom edge of the sheet to the fold and flap edge respectively. θ is the angle of tearing. The zoomed in view on the right side shows a section of the flap and the action of the stretching stress in the flap.

First, we seek a relation between stretching force in the flap and adhered sheet and adhesion strength to the substrate. The total energy of the system can be split as, $U = U_E + 2\gamma t s + \tau A$, where the terms correspond to elastic energy, fracture energy and adhesion energy, respectively.

Here, t is the graphene sheet nominal thickness, s is the length of the crack, γ is the work of fracture of the film per unit cross-sectional area, and τ is the adhesive energy per unit surface area of interface, and A and 2ts are the peeling and tearing surface areas, respectively (Figure S1). The elastic energy itself can be split as, $U_E = U_{bend} + U_{stretch}$, where the two terms correspond to bending energy (associated in the bent cylindrical ridge) and stretching energy (stored in the flat portion of the flap and the adhered sheet). In the remainder of this document, the part of the sheet being torn off and bent back is referred to as the flap, and the rest of the adhered sheet is referred to as the flat sheet.

In contrast to the earlier continuum analysis,[21] from the atomistic simulations we find that the height of the fold, h, changes during the bending-dominated regime as a function of the adhesion energy, whereas it remains approximately constant as a function of adhesion energy in the stretching-dominated regime. This implies that the contribution from bending energy becomes constant once stretching dominates. As such, bending can be regarded as a constant contribution to the total stored elastic energy. This total stored elastic energy can then be written as a function of the stretch in the flap as, $U_E = U_E(x - l, W)$. In a displacement controlled experiment, the first variational of the total energy, U, with respect to the various geometric parameters is,

$$\delta U = (\partial_W U_E)_{x,l} \delta W + (\partial_l U_E)_{x,W} + 2\gamma t \delta s + \tau W \delta l .  \tag{S1}$$

In addition, the applied force is given by the work theorem as $F = (\partial_x U_E)_{l,W}$. Using this relation for the force, along with the condition for mechanical equilibrium of the system,[56] $\partial U / \partial s = 0$, from Equation (S1), we get

$$F \cos\frac{\theta}{2} = 2\gamma t + \tau W \cos\frac{\theta}{2} - 2(\partial_W U_E)_{x,l} \sin\frac{\theta}{2} .  \tag{S2}$$

To obtain the angle of crack propagation, we apply the maximum energy release-rate criterion,[56] $\partial_\theta(\delta U/\delta s) = 0$, which using Equation (S2) yields,

$$F \sin\frac{\theta}{2} = \tau W \sin\frac{\theta}{2} + 2(\partial_W U_E)_{x,l} \cos\frac{\theta}{2}. \qquad (S3)$$

These equations can be simplified to,

$$\begin{aligned} F &= \tau W + 2\gamma t \cos\frac{\theta}{2}, \\ (\partial_W U_E)_{x,l} &= \gamma t \sin\frac{\theta}{2}, \end{aligned} \qquad (S4)$$

by projection onto the X-Y coordinate system (see Figure S1).

Up to this point, the analysis is identical to that of Hamm et al[21] but now we consider the contribution due to stretching of the straight portion of the flap, and of the adhered sheet ahead of the bent edge.

The applied force can be calculated from the stretching strain in the flap, as, $F = \sigma W t = E\varepsilon W t$. When the width is large, such that $\tau W \gg \gamma t$, we get, from Equation (S4), $F \approx \tau W$, and thus,

$$\varepsilon \approx \tau/Et. \qquad (S5)$$

This predicts a linear relation between strain in the flap, $\varepsilon$, and the adhesion energy, $\tau$. Both of these quantities can be easily measured in the atomistic simulations and are plotted in Figure S2. The value of $1/Et$ for graphene modeled by ReaxFF[30] is 0.0025 m² J⁻¹. A linear square fit to the $\varepsilon(\tau)$ numerical data yields a slope of 0.0018±0.00005 m² J⁻¹, which is consistent, and of the same order of magnitude as the predicted value.

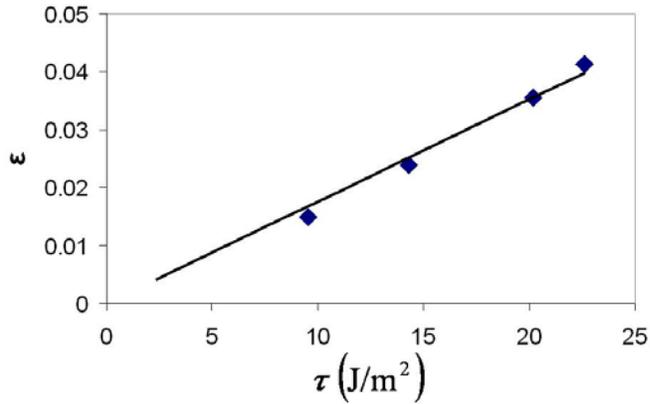

Figure S2. Tensile strain $\varepsilon_{yy}$ in the flap measured as a function of the adhesion strength from molecular dynamics experiments. Pulling speed is held constant. The slope of the linear fit is $0.0018 \pm 0.00005$ m$^2$ J$^{-1}$.

To obtain the angular dependence of the tearing on the adhesion strength, we need an estimate of the elastic energy stored both in the sheet and in the flap (second part of Equation S4), as follows. Consider an infinite isotropic material with two semi-infinite cracks parallel to each other, separated by a width W. Let the strip be under far-field uniform plane stress $\sigma_{xx} = \sigma$ (Figure S3).

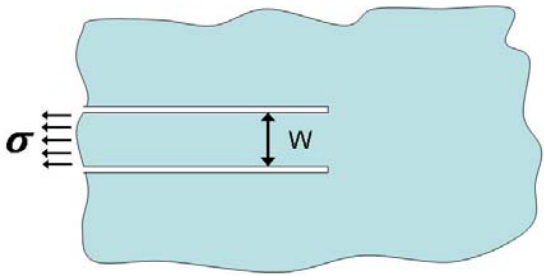

Figure S3. Schematic of an infinite sheet with two semi-infinite cracks, separated by W. The strip of material between the cracks in under a far-field tensile stress of $\sigma$.

The energy release rate under this loading predicted by energy balance using linear elastic fracture mechanics is:[57]

$$G_{stretch} = \frac{\sigma^2 W}{4E}. \tag{S6}$$

The stretching configuration in our tearing experiments resembles this geometry and loading conditions. Our system is however different from this idealized system for the following reasons: a) the finite size of the crack, and (b) the finite dimensions of the sheet. The effect of these two geometric factors, i.e. finite surface crack and finite specimen dimensions, can be grouped into a dimensionless numerical factor f, so that,

$$G_{stretch} = f \frac{\sigma^2 W}{4E}. \tag{S7}$$

This is the energy release rate per unit thickness and per unit length advance of the crack and by integrating over the crack length and thickness of sheet, we obtain,

$$U_{stretch} = f \frac{\sigma^2 W}{4E} lt + U_{stretch}^{no\,crack}(W), \tag{S8}$$

where $U_{stretch}^{no\,crack}(W)$ is the stretching energy stored in the sheet under external load when no crack is present, and is a function of W alone.

To calculate the second term in Equation (S8), we consider a semi-infinite thin plate under an edge load of magnitude $\sigma W t$, as shown schematically in Figure S4.

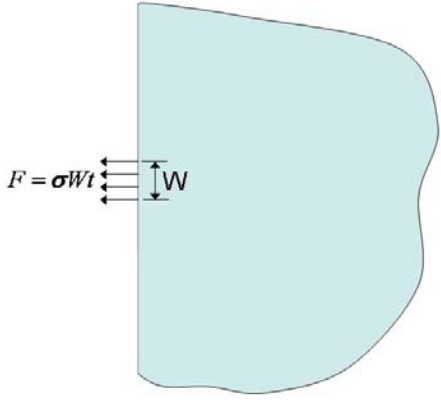

Figure S4. Schematic of a semi-infinite thin plate under a load F spread over a width W on the edge.

For points on the plate far away from the location of the applied load, this load can be considered as a point force. In that case, according to linear elasticity theory[52] the displacement field far away from the load application is,

$$u \propto \frac{F}{Et} = \frac{\sigma W}{E}. \tag{S9}$$

Hence, the strain energy stored in the plate scales as $U_{stretch}^{nocrack} = \beta \frac{\sigma^2 W^2 t}{E}$, where $\beta$ is a dimensionless numerical prefactor. Thus, the total stretching energy, from Equation (S8), is,

$$U_{stretch} = f \frac{\sigma^2 W}{4E} lt + \beta \frac{\sigma^2 W^2 t}{E}. \tag{S10}$$

The relation between the applied far-field stress and the adhesion energy (Equation S5) can now be substituted into Equation (S10) to obtain the adhesion strength dependence of the elastic stretching energy as,

$$U_{stretch} = \left( f \frac{Wlt}{4} + \beta W^2 t \right) \frac{\tau^2}{Et^2}. \tag{S11}$$

Finally, when the contribution from stretching energy dominates the elastic energy in the sheet, the second part of Equation (S4) becomes,

$$\gamma t \sin\frac{\theta}{2} = (\partial_W U_E)_{x,l} \approx (\partial_W U_{stretch})_{x,l}, \tag{S12}$$

and after rearrangement, we arrive to an expression that relates the sine of half of the tearing angle with the various geometric and material properties of the system:

Or,

$$\sin\left(\frac{\theta}{2}\right) \approx \left(\frac{fl}{4} + 2\beta W\right)\frac{\tau^2}{\gamma E t^2}. \tag{S13}$$

After calculating mean values we obtain,

$$\left\langle \sin\left(\frac{\theta}{2}\right)\right\rangle \approx \left(\frac{f\bar{l}}{4} + 2\beta\overline{W}\right)\frac{\tau^2}{\gamma E t^2} \propto \tau^2, \tag{S14}$$

where $\left\langle \sin\left(\frac{\theta}{2}\right)\right\rangle$, $\bar{l}$ and $\overline{W}$ are mean values of sine of half of the tearing angle, the crack length and the width of flap over the simulation, respectively.

We have thus rationalized the two tearing regimes present in the data presented in Figure 3a of the main text. The first regime (at low adhesion strengths, $2 < \tau < 8$ J m$^{-2}$) is characterized by the scaling, $\sin\frac{\theta}{2} \propto \sqrt{\tau}$, and the contribution to the elastic energy is dominated by bending. In the second regime (at high adhesion strengths, $\tau > 8$ J m$^{-2}$), stretching dominates and the new scaling we arrived at in Equation (S14), $\sin\frac{\theta}{2} \propto \tau^2$, is observed.

**Deformation fields as a function of adhesion energy**

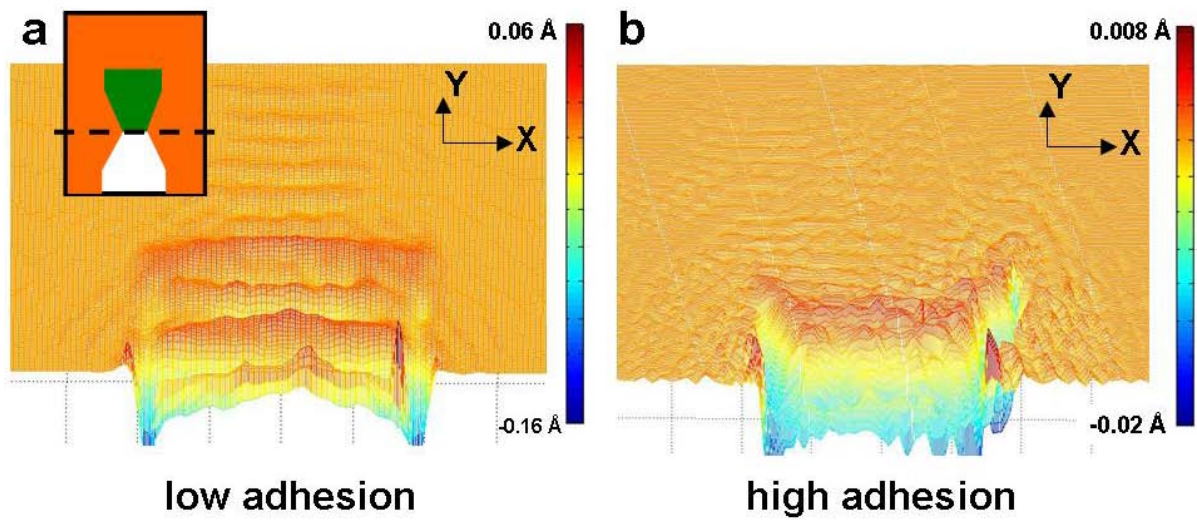

Figure S5. Surface plots of the out-of-plane (Z) deformation fields in the graphene sheet for (a) 'low' and (b) 'high' adhesion cases, as shown in Figure 4 in the main text, are calculated from molecular dynamics simulations. The inset in (a) shows a dotted line along which the view is cut off for ease of visualization and the viewing angle of the plots is in the XZ plane, at an angle to the Z axis. Out-of-plane displacements in the sheet ahead of the bent edge are seen for both cases, and the maximum gradients in the displacement (indicating local geometric torsion in the sheet) are seen to correspond to the high $\sigma_{ZX}$ shear stress regions seen in the stress plots.

**Graphene oxide**

We have studied the flakes of graphene oxide, deposited on silicon oxide substrate. No tears, similar to those observed in graphene were found. Although the method for preparation and deposition of graphene oxide flakes (exfoliation of graphite oxide by agitation in water with subsequent spin-casting on the silicon oxide surface) does not necessarily facilitate the production of such tears (unlike that in graphene, which has been obtained by the micromechanical exfoliation), we suspect that the reason might be more fundamental and related to the very different mechanical properties of graphene oxide, and the defect and functionalization structure in the sheet.

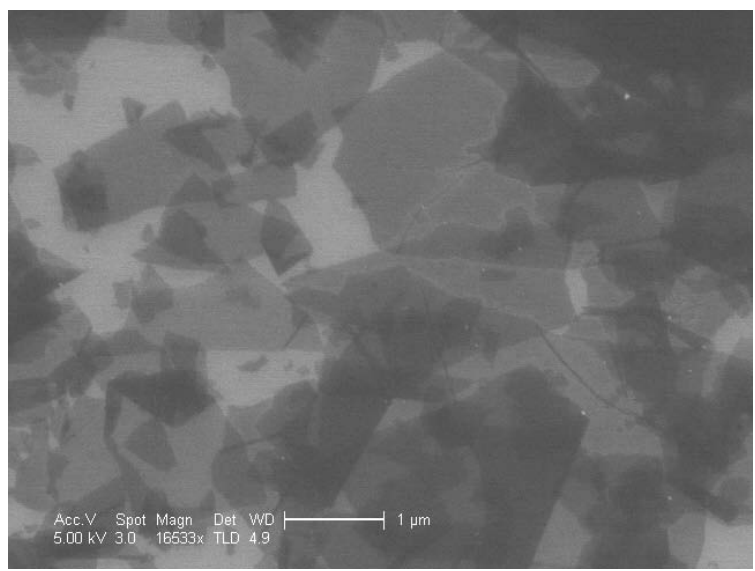

Figure S6. Flakes of graphene oxide as deposited from suspension on silicon oxide substrate by spin coating (SEM image). No tears, similar to those in graphene, have been observed.

**Supplementary References**

[58]   B. R. Lawn, Fracture of Brittle Solids, Cambridge Univ. Press, 1993.

[59]   Z. Suo, Applied Mechanics Reviews 1990, 43, S276.